\documentclass{nature}
\usepackage{amssymb}
\usepackage[utf8]{inputenc}
\usepackage[]{graphicx,bm,color,subfigure,amsmath}
\usepackage{graphicx}
\usepackage{caption}

\usepackage{savesym}
\savesymbol{spacing}

\usepackage{setspace}

\usepackage[export]{adjustbox}
\usepackage[utf8]{inputenc}
\usepackage[unicode=true]{hyperref}
\usepackage{hyperref}
\usepackage{upgreek}

\newcommand{\beginsupplement}{%
        \setcounter{table}{0}
        \renewcommand{\thetable}{S\arabic{table}}%
        \setcounter{figure}{0}
        \renewcommand{\thefigure}{S\arabic{figure}}%
     }

\makeatletter
\let\saved@includegraphics\includegraphics
\AtBeginDocument{\let\includegraphics\saved@includegraphics}
\renewenvironment*{figure}{\@float{figure}}{\end@float}
\makeatother

\title{Freezing and thawing magnetic droplet solitons 
}

\author{Martina~Ahlberg$^{1,\ast}$, Sunjae~Chung$^{1,3,\ast,+}$, Sheng~Jiang$^{1,4,5}$, Q.~Tuan~Le$^{1, 5}$, Roman Khymyn$^{1}$, Hamid Mazraati$^{2,5}$, Markus Weigand$^6$, Iuliia Bykova$^6$, Felix Gro{\ss}$^6$, Eberhard Goering$^6$, Gisela Sch\"{u}tz$^6$, Joachim Gr\"{a}fe$^6$, \& Johan~\AA{}kerman$^{1,2,5,+}$}

\begin{document}

\maketitle

\begin{affiliations}
 \item Department of Physics, University of Gothenburg, 412 96 Gothenburg, Sweden
 \item NanOsc AB, 164 40 Kista, Sweden
 \item Department of Physics Education, Korea National University of Education, Cheongju 28173, Korea
  \item School of Microelectronics, Northwestern Polytechnical University, Xi’an 710072, China
 \item Department of Applied Physics, School of Engineering Sciences, KTH Royal Institute of Technology, 100 44 Stockholm, Sweden
  \item Max Planck Institute for Intelligent Systems, Stuttgart, Germany
 
$^\ast$These authors contributed equally to this work.
\end{affiliations}

\begin{abstract}

Magnetic droplets are non-topological magnetodynamical solitons displaying a wide range of complex dynamic phenomena with potential for microwave signal generation. Bubbles, on the other hand, are internally static cylindrical magnetic domains, stabilized by external fields and magnetostatic interactions. In its original theory, the droplet was described as an imminently collapsing bubble stabilized by spin transfer torque and, in its zero-frequency limit, as equivalent to a bubble. Without nanoscale lateral confinement, pinning, or an external applied field, such a nanobubble is unstable, and should collapse. Here, we show that we can freeze dynamic droplets into static nanobubbles by decreasing the magnetic field. While the bubble has virtually the same resistance as the droplet, all signs of low-frequency microwave noise disappear. The transition is fully reversible and the bubble can be thawed back into a droplet if the magnetic field is increased under current. Whereas the droplet collapses without a sustaining current, the bubble is highly stable and remains intact for days without external drive. Electrical measurements are complemented by direct observation using scanning transmission x-ray microscopy, which corroborates the analysis and confirms that the bubble is stabilized by pinning. 
\end{abstract}

\flushbottom
\maketitle

\thispagestyle{empty}

\begin{figure}
\includegraphics [width=4in]{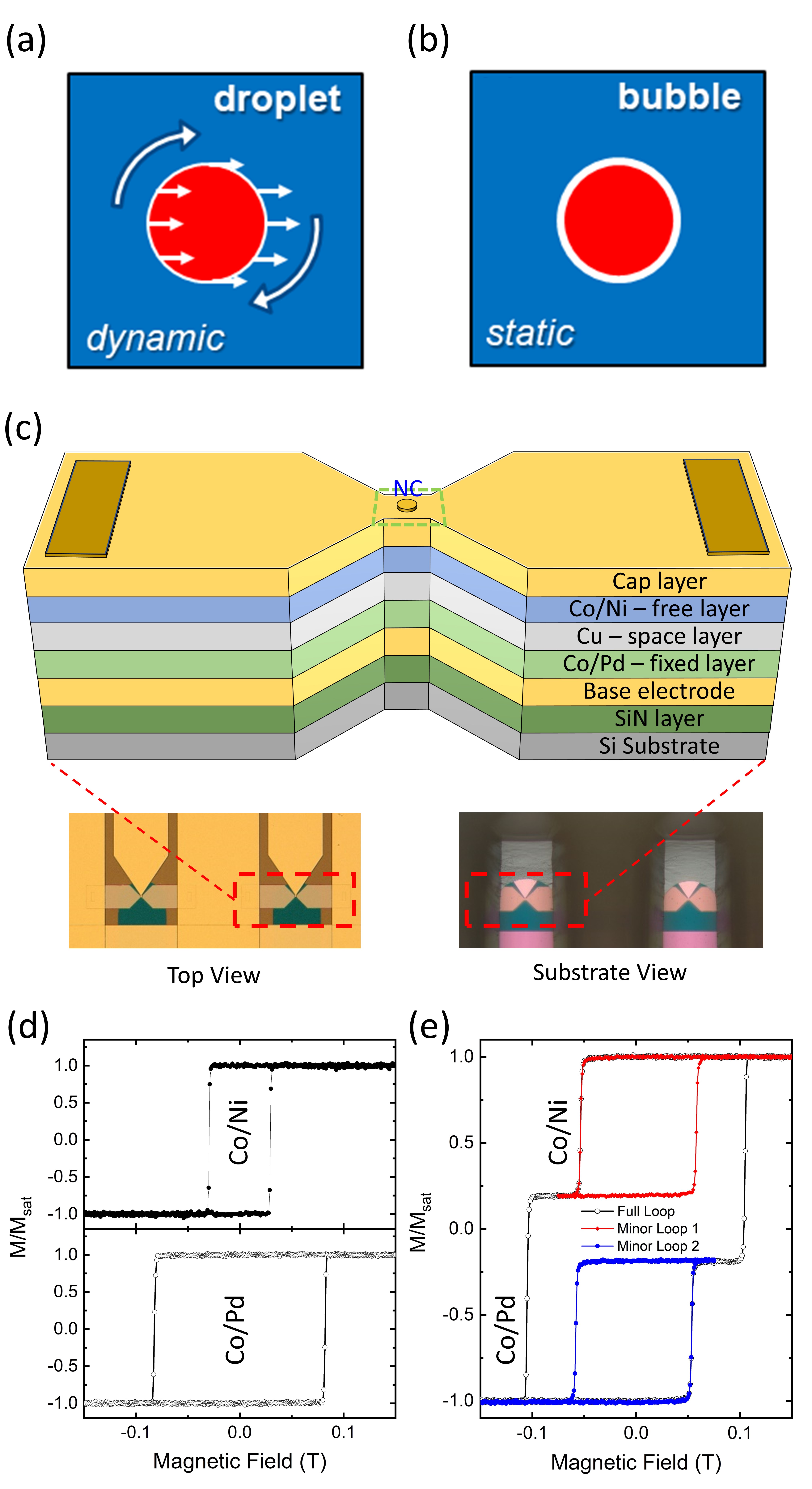}
\centering
\caption{ \label{figure1} \textbf{Droplet vs.~bubble, device structure and layout, and magnetic characterization.
} (\textbf{a}) Schematic of dynamical magnetic droplet soliton. (\textbf{b}) Schematic of a static magnetic bubble. (\textbf{c}) Schematic of an all-perpendicular STNO composed of [Co/Pd] (fixed) and [Co/Ni] (free) multilayers with a Cu spacer fabricated on a SiN membrane structure. The insets underneath show optical micrographs of the SiN membrane areas through which the different metal layers of the device can be seen. (\textbf{d}) Hysteresis loops of single  Co/Pd and Co/Ni layers. (\textbf{e}) Hysteresis loop of a full [Co/Pd]/Cu/[Co/Ni] stack. 
}
\end{figure}

Magnetic droplets are intrinsically dynamic, non-topological, magnetodynamical solitons\cite{Ivanov1977, Hoefer2010,Mohseni2013, Mohseni2013b,Iacocca2014, Macia2014, Chung2014, Lendinez2015, Chung2015, Chung2016, Xiao2017prb, Lendinez2017prapplied, Sulymenko2018, Statuto2019, Mohseni2020PRB}, which can be nucleated and sustained both in spin torque nano-oscillators (STNOs)\cite{Mohseni2013, Macia2014, Lendinez2015, Lendinez2017prapplied, Statuto2019} and spin Hall nano-oscillators (SHNOs)\cite{Divinskiy2017}, provided the magnetodynamically active layer has sufficient perpendicular magnetic anisotropy (PMA). Magnetic droplets are characterized by a reversed core separated from the surrounding magnetization via a perimeter of precessing spins (See Fig.1(a))\cite{Hoefer2010, Mohseni2013}. While first predicted over 40 years ago in an ideal zero-damping medium\cite{Ivanov1977}, their possible experimental realization was later suggested theoretically\cite{Hoefer2010} in STNOs with PMA free layers\cite{Mohseni2011, Rippard2010prb}. 
After the first experimental demonstration of magnetic droplets, reported in STNOs with a PMA Co/Ni free layer and a Co fixed layer\cite{Mohseni2013}, interest in magnetic droplets continues to increase due to its interesting characteristics, such as a highly nonlinear dynamics\cite{Hoefer2010, Bookman2013, Xiao2017prb}, large power emission\cite{Mohseni2013, Locatelli2014, Chung2016, Chung2018}, and possible applications in microwave-assisted magnetic recording (MAMR)\cite{Okamoto2015, Bosu2017} and neuromorphic chips as nonlinear oscillators\cite{Torrejon2017Nature, Romera2018, Macia2020}. Several theoretical\cite{Bookman2013, Maiden2014, Iacocca2014, Puliafito2014, Xiao2017prb, Wang2017, Mohseni2018, Mohseni2020PRB, Mohseni2020PRApplied, Sisodia2021, Yazdi2021} and experimental\cite{Macia2014,Chung2014, Chung2015, Lendinez2015, Backes2015, Carpentieri2015, Chung2016, Lendinez2017prapplied, Divinskiy2017, Hang2018, Chung2018, Jiang2018IEEEML, Shi2020} studies on magnetic droplets have since been presented.

As pointed out by Hoefer \emph{et al.}, the droplet is reminiscent of a magnetic bubble\cite{Hoefer2010} (Fig.1(b)) and they identify a possible zero-frequency droplet with a topologically trivial magnetic bubble\cite{Nielsen1976, Giess1980, DeLeeuw1980, Komineas1996, Moutafis2009}. Despite the large number of experimental droplet studies, the low-field/low-frequency behavior of droplets has not yet been explored and the relation between droplets and bubbles --- as well as a possible transition between the two --- remain unclear. In order to explore these phenomena, we here study magnetic droplets specifically in the low-field regime using both electrical and microwave spectroscopy measurements as well as the direct microscopical observation based on Scanning Transmission X-ray Microscopy (STXM). We find clear experimental evidence for a droplet-to-bubble transition as the field strength, and hence the droplet frequency, is reduced, and a reversible bubble-to-droplet transition as the field is again increased in an attempt to squash the bubble, provided stabilizing spin transfer torque is still present via the STNO current. Our experimental results hence corroborate the picture, first expressed by Hoefer \emph{et al.}, that a magnetic droplet can be viewed  \textquotedblleft\emph{as an imminently collapsing bubble that is critically stabilized by the localized injection of spin torque}\textquotedblright.

Figure~\ref{figure1} shows a schematic of the studied all-perpendicular STNOs, comprised of a
[Co/Pd]/ Cu/[Co/Ni] GMR stack deposited on a Si${_3}$N${_4}$ membrane (for fabrication details, please see Methods). Underneath the schematic we show two optical microscopy images taken from opposite directions to highlight the optical transmission of the Si${_3}$N${_4}$ membrane. In Fig.~\ref{figure1}(d) we show the magnetic properties of the individual free and fixed layers based on calibration samples, and their combined behavior in full STNO stacks in Fig.~\ref{figure1}(e).

\begin{figure}
\includegraphics [width=0.64\textwidth]{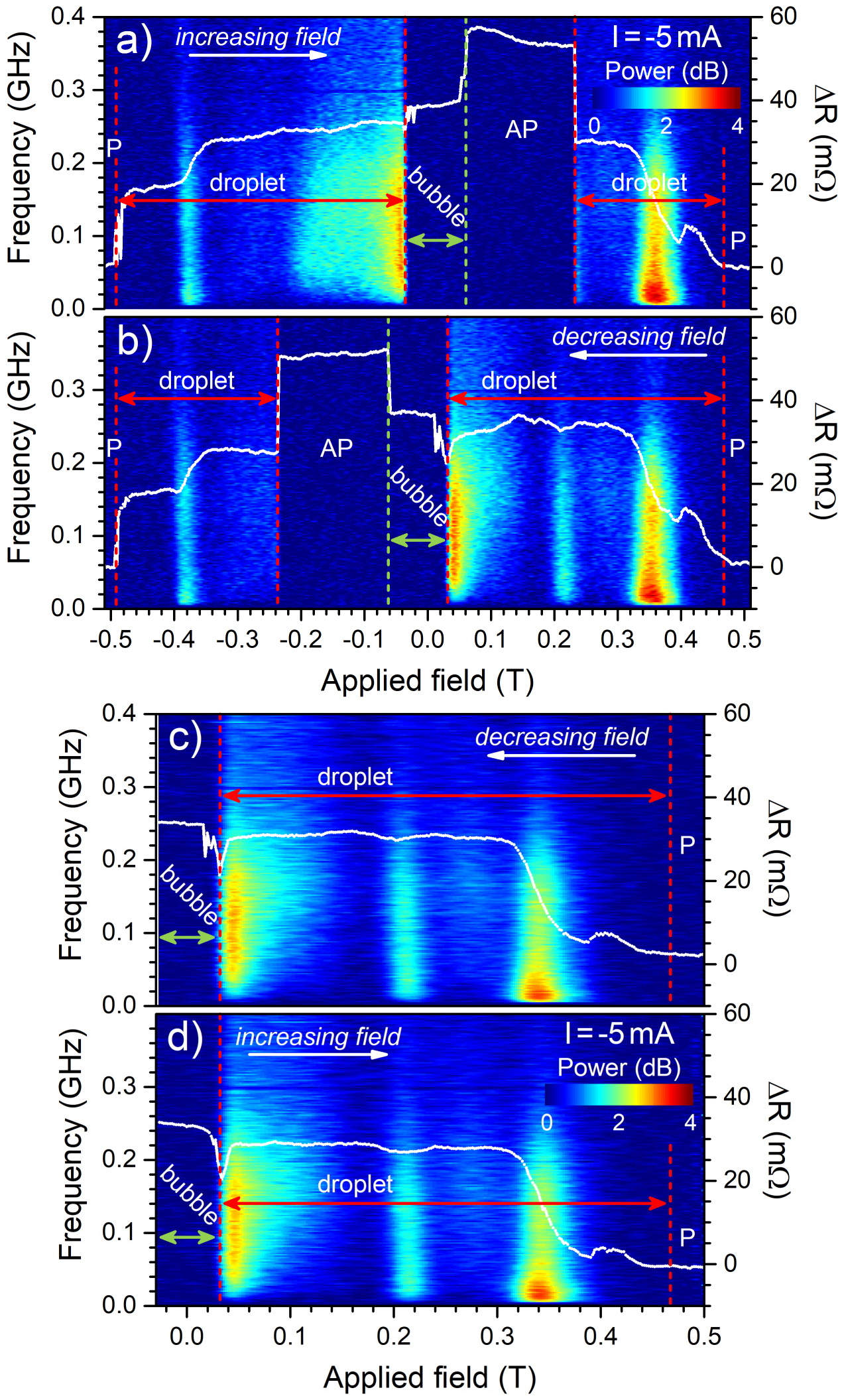}
\centering
\caption{ \label{figure2} \textbf{Microwave noise and STNO resistance vs.~field.} (\textbf{a})-(\textbf{d}) Color plot of the power spectral density (PSD) of the microwave noise as a function of decreasing (\textbf{a},\textbf{c}) and increasing (\textbf{b},\textbf{d}) field, with the STNO resistance (white line) overlayed; the applied current is $-5$~mA. (\textbf{a},\textbf{b}) Wide field sweep covering full saturation at both positive and negative fields. P/AP indicate the parallel/antiparallel state of the STNO; red arrow indicates the droplet region, and green arrow the bubble region. (\textbf{c},\textbf{d}) Minor field sweeps showing how the droplet/bubble transition is fully reversible.
}
\end{figure}

Figure~\ref{figure2} presents the resistance and microwave signal as a function of field for an applied current of --5~mA. The field is first increased from --0.51~T to 0.51~T in Fig.~\ref{figure2}(a) and then decreased from positive to negative field in Fig.~\ref{figure2}(b). At large negative fields, the STNO is in its lowest resistance state, consistent with a parallel (P) relative orientation of its free and fixed layers. At about \nobreakdash--0.49~T, the resistance increases about 20 mOhm in a step-like fashion and there is a slight increase in the microwave noise background, both strong indications of the nucleation of a droplet. At about \nobreakdash--0.38~T, there is a second step-like increase in the STNO resistance and a marked further increase in the microwave noise. We interpret this as a transition into a larger droplet as the opposing applied field is reduced. At yet lower fields the droplet continues to grow in size (the STNO resistance increases), while its stability seems to deteriorate as indicated by the growing intensity of the microwave noise background. At about --0.04~T, the microwave noise rapidly reaches a maximum and then suddenly disappears altogether, while the resistance exhibits a small jump of about 5 mOhm. The complete microwave silence indicates that the magnetic state is now static, and we are lead to conclude that the droplet precession has stopped entirely and that the droplet has transitioned into a nanobubble state.

The nanobubble resistance exhibits jumps reminiscent of Barkhausen noise\cite{Kim2003PRL, Balk2014PRB, Herranen2019PRL}, indicating pinning possibly at grain boundaries or defects of the sputtered film. When the field is further increased, the bubble resistance increases gradually, indicating a continued growth of its size. At about 0.06~T, the entire free layer switches its magnetization direction and the antiparallel (AP) state is clearly identifiable in the resistance. When the fixed layer switches at 0.23~T, a droplet is immediately nucleated. With further increasing of the opposing field, the droplet again shows a gradual transition to a smaller size; the droplet finally disappears as the STNO transitions into a full P state at about 0.47~T. The overall behavior is very similar for decreasing fields (Fig.2b) where the same P/AP/droplet/nanobubble states can be clearly identified via the STNO resistance and the microwave noise.

As mentioned above, the microwave noise power is far from constant for the whole droplet region and peaks at certain fields. We identify these peaks as marks of mode hopping between different droplet states\cite{Statuto2019}. While the details of the spectrum is highly reproducible (cf. increasing and decreasing fields) and serves as a fingerprint for each device, the patterns at negative and positive fields are quite different. The magnetoresistance implies that a relatively small and stable droplet ($\mu_{0}H < -0.4$~T) is abruptly followed by a larger but similarly stable mode. In contrast to the symmetric noise patterns around the droplet-to-droplet transitions, the strong increase in microwave noise power around the droplet-to-bubble transition is highly asymmetric. There is first an extended field region of monotonic increase in the noise, which is then abruptly cut off and replaced by a completely silent bubble state. This highlights the very different non-dynamical nature of the nanobubble and suggests that mode hopping out of the nanobubble state and back into a droplet state is negligible, once the nanobubble has formed. 

Figure~\ref{figure2}(c) and (d) demonstrate that it is possible to freeze the dynamic droplet into a static bubble and then thaw it back into a droplet using only the magnetic field under constant spin transfer torque. In particular, Fig.~\ref{figure2}(d) shows how the nanobubble first is about to collapse at 0.025~T as it is getting squeezed by the opposing pressure from the increasing applied field. There is some slight Barkhausen noise in the rapidly dropping resistance, but otherwise no measurable microwave noise. However, instead of switching to a P state, the resistance then exhibits a sharp minimum after which it shows a rapid increase, which is accompanied by a high level of microwave noise. The collapsing nanobubble is hence rescued by the stabilizing spin transfer torque, which sets the spins in the bubble perimeter into precessional motion and restores the full dynamics of a magnetic droplet. Judging from the resistance, it is noteworthy that the droplet is slightly larger than the smallest nanobubble. Within the experimental accuracy (a field step of 2~mT), we do not observe any significant hysteresis in this transition. Hence there is a negligible energy barrier between the two states and the bubble can indeed be viewed as a zero-frequency droplet, albeit still likely affected by pinning.

\begin{figure}
\includegraphics [width=1\textwidth]{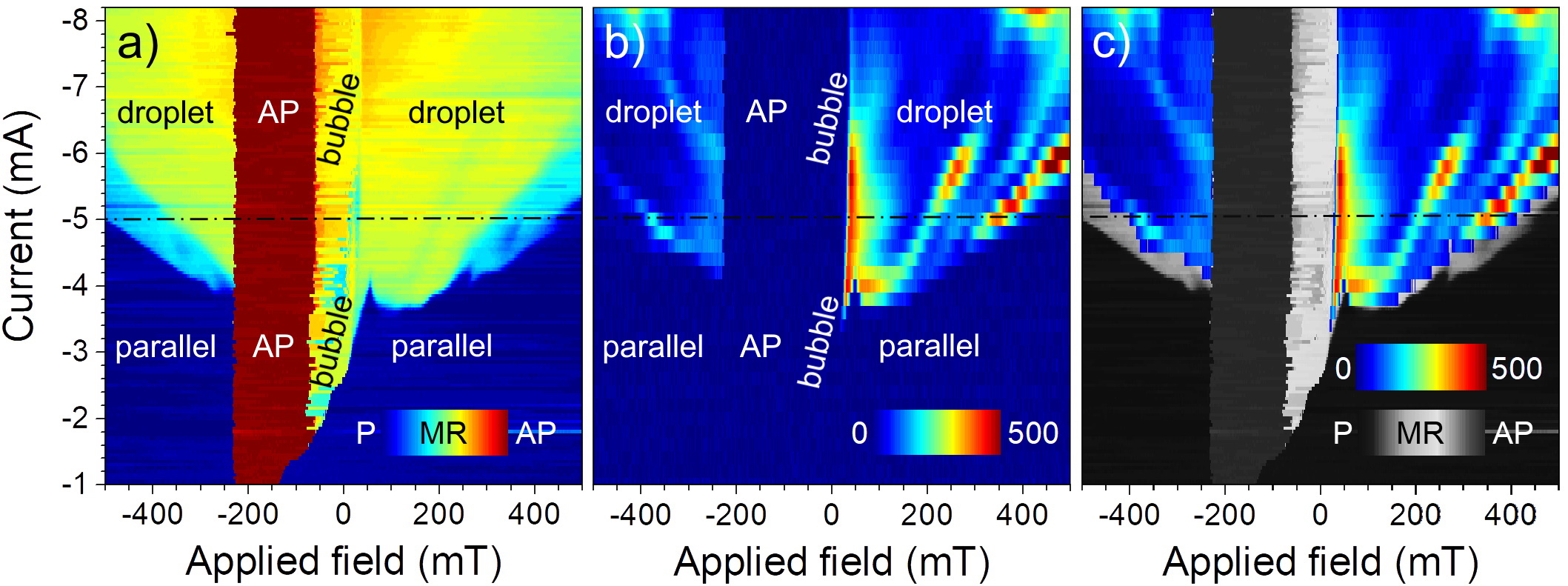}
\centering
\caption{ \label{figure3} \textbf{Phase diagrams based on the resistance and the microwave 
noise.} (\textbf{a}) STNO resistance and (\textbf{b}) integrated (0--0.5~GHz) microwave noise level as a function of field and current. (\textbf{c}) shows the noise level in (\textbf{b}) overlaid on the resistance in (\textbf{a}) displayed using a gray scale highlighting intermediate resistance levels indicative of droplets/bubbles. The dash-dotted black line corresponds to the field-sweep at $I$ = --5~mA given in Fig.~\ref{figure2}. The parallel (P) and antiparallel (AP) states are easily discernible in the MR-map (\textbf{a}) as dark blue and dark red, while both the droplet and the bubble are characterized by intermediate resistance in green--yellow. The stark difference between the droplet and the bubble is revealed in the noise spectrum (\textbf{b}), where the stability of the bubble is manifested. Note however that the light-blue flanges in (\textbf{a}) correspond to a different droplet regime not captured in the microwave signal presented in (\textbf{b}).
}
\end{figure}

Figure~\ref{figure3}(a) presents a phase diagram based on a two-dimensional map of the STNO resistance as functions of current and field. All data was acquired in a decreasing field at a constant current level. The parallel (P) and antiparallel (AP) configurations are easily identified by the dark blue and dark red colors, respectively, and for current magnitudes below 1.8~mA, these are the only two available states, as expected for a GMR device. However, even at these weak currents, the P$\rightarrow$AP switching field  is  clearly affected by the STT from the nanocontact; in contrast, the AP$\rightarrow$P switching field is entirely unaffected. In an intermediate current region, from about --1.8 to --3.5~mA, the STT can not yet sustain a droplet but is sufficient to create a nanobubble directly from the P state. As magnetic switching typically involves both domain nucleation and domain propagation, we interpret this current dependent switching in the following way (see Supplementary Materials for a zoom-in of this particular part of the phase diagram). For current magnitudes below 1~mA, magnetic switching is limited by the field required for domain nucleation and, in addition, the location of initial domain nucleation is far from the nanocontact region as STT from the current has no discernible impact. However, for current magnitudes above 1~mA, where we observe a strong current dependence of the switching field, we conclude that the domain nucleation has moved to underneath the nanocontact. If we reduce the field magnitude, we need a stronger current to assist in the domain nucleation, but once formed, it propagates through the entire free layer. However, at fields weaker than the field needed for domain propagation, i.e. the pinning field, which we read out as about 60 mT, the nucleated domain is no longer able to propagate and instead remains as a nanobubble directly underneath the nanocontact. The nanobubble can hence form either from the P state or from a droplet.

The droplet shows two discernable states, a high-field/low-current mode that exhibits a rather small MR (light blue). This mode moves to higher fields with increasing current and is no longer visible above $\approx -6$~mA. The other distinguishable droplet mode is characterized by an intermediate resistance (green-yellow). The bubble is almost indiscernible from the latter droplet state, even though a subtle line traces out the transition between the two. Moreover, the bubble resistance is not a smooth function of applied field, but displays notches and steps, indicative of Barkhausen noise due to pinning. 

In contrast to their almost identical resistance, a stark difference between the droplet and the bubble is uncovered in Fig.~\ref{figure3}(b), where we show the microwave signal integrated over 0--0.5~GHz. The droplet exhibits non-zero power levels of low frequency microwave noise, while the P, AP, and bubble states are definitely static and silent. Figure~\ref{figure3}(b) also further unveils the complex relation between the applied field and current, and the particular droplet characteristics. A strong microwave noise signal denotes mode hopping and these events exhibit a strong dependence on both  field and current. We can identify three traces of mode hopping for positive fields, while there is only two weak trails at negative fields. There is also regions where the droplet is very stable and the noise level is almost zero. These features act like fingerprints for each measured device and are highly reproducible in consecutive measurements, but differ between STNOs. We then overlay the microwave noise data onto the resistance data, now plotted with a gray scale that highlights intermediate resistance levels (Fig.~\ref{figure3}(c)). Parts of the low-field/low-current droplet regime (light blue in Fig.~\ref{figure3}(a)) does not exhibit any measurable microwave noise. It is possible that its dynamics is on a slower time scale than the microwave frequencies our set-up is sensitive to.

\begin{figure}
\includegraphics [width=4.6in]{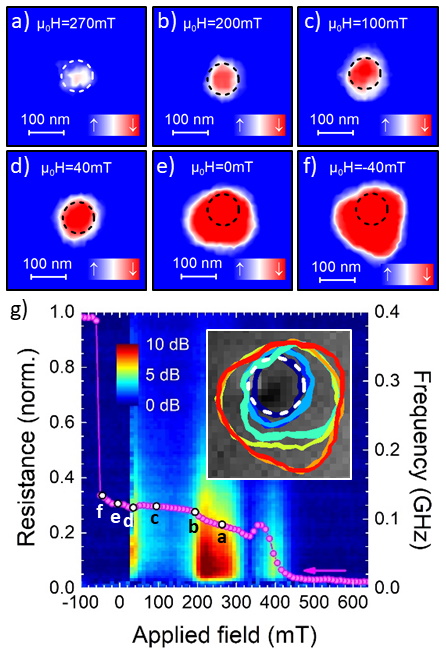}
\centering
\caption{ \label{figure4} \textbf{Scanning Transmission X-ray Microscopy (STXM).} (\textbf{a})--(\textbf{f}) STXM images of the nanocontact region vs.~decreasing field for a current of -7~mA. Blue corresponds to magnetization alligned with the applied field, red corresponds to magnetization anti-alligned with the applied field, whereas white indicates in-plane magnetization. The STNO resistance and the microwave noise PSD vs.~decreasing field are shown in (\textbf{g}) where the points corresponding to the STXM images have been labelled \textbf{a}--\textbf{f}. The inset in (\textbf{g}) highlights the perimeter of the droplet/bubble as the applied field is decreased from 200~mT (dark blue) to 60~mT (blue), and further reduced to -40~mT (red) in steps of 20~mT.
}
\end{figure}

We finally turn to the results of the scanning transmission X-ray microscopy results, illustrated in Fig.~\ref{figure4}. Images of the droplet/bubble are shown in Fig.~\ref{figure4}(a)-(f), and the corresponding magnetoresistance and microwave signal are presented in Fig.~\ref{figure4}(g) with the matching field of the images marked by their letter. The STXM and the electrical measurements were performed in separate setups, hence there is a small uncertainty in comparing the field values of the two, although both measurements seem highly consistent with each other. The dashed white or black circles mark the position of the nanocontact. It has been placed by assuming that the droplet/bubble in Fig.~\ref{figure4}(d) is centered under the NC and by comparing the non-magnetic contrast of the different images. The method works very well as confirmed by the good overlap of the perimeters in the inset of Fig.~\ref{figure4}(g), but it should be remembered that the absolute position is still based on this assumption. 

Figure~\ref{figure4}(a) is measured at 270~mT, and shows a mode associated with a high noise level in Fig.~\ref{figure4}(g). Only a weak and mostly white feature is captured in the STXM image. STXM measures a time averaged image and the droplet is in this highly noisy regime expected to experience large drift instabilities and continuously vanish and renucleate underneath the nanocontact. As a consequence, only a washed out and poorly reversed feature results. In contrast, Fig.~\ref{figure4}(b)-(d) display more stable and more clearly reversed droplets.  
They have approximately the same radius as the nanocontact, although the size definitely increases slightly with decreasing field, as expected. We have in  earlier STXM work observed a significant effect of the Zhang-Li torque on the droplet size\cite{Chung2018, Albert2020}. The magnitude of this effect depends on the current density ($j_\mathrm{dc}$) and we have performed simulations which confirm that the difference between the droplet diameter here and in our former publication is indeed due to a weaker $j_\mathrm{dc}$. At zero field, a bubble is clearly formed and it prevails down to -40~mT (Fig.~\ref{figure4}(e)-(f)). It is no longer centered on the nanocontact, but has mostly expanded in one direction. It should be noted though, that the field in the microscope is given by rotating permanent magnets and the sample may have been subjected to in-plane fields between two set values. Nevertheless, the inset in Fig.~\ref{figure4}(g) presents the perimeter of the droplet/bubble as the field decreases from 200~mT (dark blue) to -40~mT (red), and the initial bubble at 40~mT (light blue) grows in distinct steps, which implies that the size is controlled by pinning.

Returning to the original droplet theory of Hoefer et al, we note that pinning was not included.\cite{Hoefer2010} It is clear from our experimental observations that pinning has a strong and immediate impact on the relation between droplets and nanobubbles and must be included in the low-field/low-current regime. Instead of exhibiting a continuous slow-down and frequency decrease to zero with decreasing field, there is a minimum droplet precession frequency that spin transfer torque can sustain before pinning overcomes the precession. As this minimum frequency is approached from above, the broad-band microwave noise diverges as the competition between the inertia of the precession and the pinning makes the droplet dynamics increasingly erratic until pinning finally gets complete control of the precession abruptly stops, leaving complete microwave silence in its wake.


\begin{thebibliography}{10}
\expandafter\ifx\csname url\endcsname\relax
  \def\url#1{\texttt{#1}}\fi
\expandafter\ifx\csname urlprefix\endcsname\relax\def\urlprefix{URL }\fi
\providecommand{\bibinfo}[2]{#2}
\providecommand{\eprint}[2][]{\url{#2}}

\bibitem{Ivanov1977}
\bibinfo{author}{Ivanov, B.} \& \bibinfo{author}{Kosevich, A.}
\newblock \bibinfo{title}{Bound states of a large number of magnons in a
  ferromagnet with a single-ion anisotropy}.
\newblock \emph{\bibinfo{journal}{Zh. Eksp. Teor. Fiz.}}
  \textbf{\bibinfo{volume}{72}}, \bibinfo{pages}{2000} (\bibinfo{year}{1977}).

\bibitem{Hoefer2010}
\bibinfo{author}{Hoefer, M.~A.}, \bibinfo{author}{Silva, T.~J.} \&
  \bibinfo{author}{Keller, M.~W.}
\newblock \bibinfo{title}{{Theory for a dissipative droplet soliton excited by
  a spin torque nanocontact}}.
\newblock \emph{\bibinfo{journal}{Phys. Rev. B}} \textbf{\bibinfo{volume}{82}},
  \bibinfo{pages}{054432} (\bibinfo{year}{2010}).

\bibitem{Mohseni2013}
\bibinfo{author}{Mohseni, S.~M.} \emph{et~al.}
\newblock \bibinfo{title}{{Spin Torque-Generated Magnetic Droplet Solitons}}.
\newblock \emph{\bibinfo{journal}{Science}} \textbf{\bibinfo{volume}{339}},
  \bibinfo{pages}{1295--1298} (\bibinfo{year}{2013}).

\bibitem{Mohseni2013b}
\bibinfo{author}{Mohseni, S.} \emph{et~al.}
\newblock \bibinfo{title}{Magnetic droplet solitons in orthogonal nano-contact
  spin torque oscillators}.
\newblock \emph{\bibinfo{journal}{Physica B}} \textbf{\bibinfo{volume}{435}},
  \bibinfo{pages}{84 -- 87} (\bibinfo{year}{2014}).

\bibitem{Iacocca2014}
\bibinfo{author}{Iacocca, E.} \emph{et~al.}
\newblock \bibinfo{title}{Confined dissipative droplet solitons in spin-valve
  nanowires with perpendicular magnetic anisotropy}.
\newblock \emph{\bibinfo{journal}{Phys. Rev. Lett.}}
  \textbf{\bibinfo{volume}{112}}, \bibinfo{pages}{047201}
  (\bibinfo{year}{2014}).

\bibitem{Macia2014}
\bibinfo{author}{Maci{\`{a}}, F.}, \bibinfo{author}{Backes, D.} \&
  \bibinfo{author}{Kent, A.~D.}
\newblock \bibinfo{title}{{Stable magnetic droplet solitons in spin-transfer
  nanocontacts}}.
\newblock \emph{\bibinfo{journal}{Nat. Nanotechnol}}
  \textbf{\bibinfo{volume}{9}}, \bibinfo{pages}{992--996}
  (\bibinfo{year}{2014}).

\bibitem{Chung2014}
\bibinfo{author}{Chung, S.} \emph{et~al.}
\newblock \bibinfo{title}{Spin transfer torque generated magnetic droplet
  solitons (invited)}.
\newblock \emph{\bibinfo{journal}{J. Appl. Phys.}}
  \textbf{\bibinfo{volume}{115}}, \bibinfo{pages}{172612}
  (\bibinfo{year}{2014}).

\bibitem{Lendinez2015}
\bibinfo{author}{Lend{\'{i}}nez, S.}, \bibinfo{author}{Statuto, N.},
  \bibinfo{author}{Backes, D.}, \bibinfo{author}{Kent, A.~D.} \&
  \bibinfo{author}{Maci{\`{a}}, F.}
\newblock \bibinfo{title}{{Observation of droplet soliton drift resonances in a
  spin-transfer-torque nanocontact to a ferromagnetic thin film}}.
\newblock \emph{\bibinfo{journal}{Phys. Rev. B}} \textbf{\bibinfo{volume}{92}},
  \bibinfo{pages}{174426} (\bibinfo{year}{2015}).

\bibitem{Chung2015}
\bibinfo{author}{Chung, S.} \emph{et~al.}
\newblock \bibinfo{title}{{Magnetic droplet solitons in orthogonal spin
  valves}}.
\newblock \emph{\bibinfo{journal}{Low Temp. Phys.}}
  \textbf{\bibinfo{volume}{41}}, \bibinfo{pages}{833--837}
  (\bibinfo{year}{2015}).

\bibitem{Chung2016}
\bibinfo{author}{Chung, S.} \emph{et~al.}
\newblock \bibinfo{title}{{Magnetic droplet nucleation boundary in orthogonal
  spin-torque nano-oscillators}}.
\newblock \emph{\bibinfo{journal}{Nat. Commun.}} \textbf{\bibinfo{volume}{7}},
  \bibinfo{pages}{11209} (\bibinfo{year}{2016}).

\bibitem{Xiao2017prb}
\bibinfo{author}{Xiao, D.} \emph{et~al.}
\newblock \bibinfo{title}{{Parametric autoexcitation of magnetic droplet
  soliton perimeter modes}}.
\newblock \emph{\bibinfo{journal}{Phys. Rev. B}} \textbf{\bibinfo{volume}{95}},
  \bibinfo{pages}{024106} (\bibinfo{year}{2017}).

\bibitem{Lendinez2017prapplied}
\bibinfo{author}{Lend{\'{i}}nez, S.} \emph{et~al.}
\newblock \bibinfo{title}{{Effect of Temperature on Magnetic Solitons Induced
  by Spin-Transfer Torque}}.
\newblock \emph{\bibinfo{journal}{Phys. Rev. Appl.}}
  \textbf{\bibinfo{volume}{7}}, \bibinfo{pages}{054027} (\bibinfo{year}{2017}).

\bibitem{Sulymenko2018}
\bibinfo{author}{Sulymenko, O.}, \bibinfo{author}{Prokopenko, O.},
  \bibinfo{author}{Tyberkevych, V.}, \bibinfo{author}{Slavin, A.} \&
  \bibinfo{author}{Serga, A.}
\newblock \bibinfo{title}{{Bullets and droplets: Two-dimensional spin-wave
  solitons in modern magnonics (Review Article)}}.
\newblock \emph{\bibinfo{journal}{Low Temp. Phys.}}
  \textbf{\bibinfo{volume}{44}}, \bibinfo{pages}{775} (\bibinfo{year}{2018}).

\bibitem{Statuto2019}
\bibinfo{author}{Statuto, N.}, \bibinfo{author}{Hahn, C.},
  \bibinfo{author}{Hern{\`{a}}ndez, J.~M.}, \bibinfo{author}{Kent, A.~D.} \&
  \bibinfo{author}{Maci{\`{a}}, F.}
\newblock \bibinfo{title}{{Multiple magnetic droplet soliton modes}}.
\newblock \emph{\bibinfo{journal}{Phys. Rev. B}} \textbf{\bibinfo{volume}{99}},
  \bibinfo{pages}{174436} (\bibinfo{year}{2019}).

\bibitem{Mohseni2020PRB}
\bibinfo{author}{Mohseni, M.} \emph{et~al.}
\newblock \bibinfo{title}{{Chiral excitations of magnetic droplet solitons
  driven by their own inertia}}.
\newblock \emph{\bibinfo{journal}{Phys. Rev. B}}
  \textbf{\bibinfo{volume}{101}}, \bibinfo{pages}{20417}
  (\bibinfo{year}{2020}).

\bibitem{Divinskiy2017}
\bibinfo{author}{Divinskiy, B.} \emph{et~al.}
\newblock \bibinfo{title}{{Magnetic droplet solitons generated by pure spin
  currents}}.
\newblock \emph{\bibinfo{journal}{Phys. Rev. B}} \textbf{\bibinfo{volume}{96}},
  \bibinfo{pages}{224419} (\bibinfo{year}{2017}).

\bibitem{Mohseni2011}
\bibinfo{author}{Mohseni, S.~M.} \emph{et~al.}
\newblock \bibinfo{title}{{High frequency operation of a spin-torque oscillator
  at low field}}.
\newblock \emph{\bibinfo{journal}{Phys. Status Solidi RRL}}
  \textbf{\bibinfo{volume}{5}}, \bibinfo{pages}{432--434}
  (\bibinfo{year}{2011}).

\bibitem{Rippard2010prb}
\bibinfo{author}{Rippard, W.~H.} \emph{et~al.}
\newblock \bibinfo{title}{{Spin-transfer dynamics in spin valves with
  out-of-plane magnetized CoNi free layers}}.
\newblock \emph{\bibinfo{journal}{Phys. Rev. B}} \textbf{\bibinfo{volume}{81}},
  \bibinfo{pages}{014426} (\bibinfo{year}{2010}).

\bibitem{Bookman2013}
\bibinfo{author}{Bookman, L.~D.} \& \bibinfo{author}{Hoefer, M.~A.}
\newblock \bibinfo{title}{Analytical theory of modulated magnetic solitons}.
\newblock \emph{\bibinfo{journal}{Phys. Rev. B}} \textbf{\bibinfo{volume}{88}},
  \bibinfo{pages}{184401} (\bibinfo{year}{2013}).

\bibitem{Locatelli2014}
\bibinfo{author}{Locatelli, N.}, \bibinfo{author}{Cros, V.} \&
  \bibinfo{author}{Grollier, J.}
\newblock \bibinfo{title}{{Spin-torque building blocks}}.
\newblock \emph{\bibinfo{journal}{Nat. Mater.}} \textbf{\bibinfo{volume}{13}},
  \bibinfo{pages}{11} (\bibinfo{year}{2014}).

\bibitem{Chung2018}
\bibinfo{author}{Chung, S.} \emph{et~al.}
\newblock \bibinfo{title}{{Direct Observation of Zhang-Li Torque Expansion of
  Magnetic Droplet Solitons}}.
\newblock \emph{\bibinfo{journal}{Phys. Rev. Lett.}}
  \textbf{\bibinfo{volume}{120}}, \bibinfo{pages}{217204}
  (\bibinfo{year}{2018}).

\bibitem{Okamoto2015}
\bibinfo{author}{Okamoto, S.}, \bibinfo{author}{Kikuchi, N.},
  \bibinfo{author}{Furuta, M.}, \bibinfo{author}{Kitakami, O.} \&
  \bibinfo{author}{Shimatsu, T.}
\newblock \bibinfo{title}{{Microwave assisted magnetic recording technologies
  and related physics}}.
\newblock \emph{\bibinfo{journal}{J. Phys. D: Appl. Phys.}}
  \textbf{\bibinfo{volume}{48}}, \bibinfo{pages}{353001}
  (\bibinfo{year}{2015}).

\bibitem{Bosu2017}
\bibinfo{author}{Bosu, S.} \emph{et~al.}
\newblock \bibinfo{title}{{High frequency out-of-plane oscillation with large
  cone angle in mag-flip spin torque oscillators for microwave assisted
  magnetic recording}}.
\newblock \emph{\bibinfo{journal}{Appl. Phys. Lett.}}
  \textbf{\bibinfo{volume}{110}}, \bibinfo{pages}{142403}
  (\bibinfo{year}{2017}).

\bibitem{Torrejon2017Nature}
\bibinfo{author}{Torrejon, J.} \emph{et~al.}
\newblock \bibinfo{title}{{Neuromorphic computing with nanoscale spintronic
  oscillators}}.
\newblock \emph{\bibinfo{journal}{Nature}} \textbf{\bibinfo{volume}{547}},
  \bibinfo{pages}{428--431} (\bibinfo{year}{2017}).

\bibitem{Romera2018}
\bibinfo{author}{Romera, M.} \emph{et~al.}
\newblock \bibinfo{title}{{Vowel recognition with four coupled spin-torque
  nano-oscillators}}.
\newblock \emph{\bibinfo{journal}{Nature}} \textbf{\bibinfo{volume}{563}},
  \bibinfo{pages}{230--234} (\bibinfo{year}{2018}).

\bibitem{Macia2020}
\bibinfo{author}{Maci{\`{a}}, F.} \& \bibinfo{author}{Kent, A.~D.}
\newblock \bibinfo{title}{{Magnetic droplet solitons}}.
\newblock \emph{\bibinfo{journal}{J. Appl. Phys.}}
  \textbf{\bibinfo{volume}{128}}, \bibinfo{pages}{100901}
  (\bibinfo{year}{2020}).

\bibitem{Maiden2014}
\bibinfo{author}{Maiden, M.~D.}, \bibinfo{author}{Bookman, L.~D.} \&
  \bibinfo{author}{Hoefer, M.~a.}
\newblock \bibinfo{title}{{Attraction, merger, reflection, and annihilation in
  magnetic droplet soliton scattering}}.
\newblock \emph{\bibinfo{journal}{Phys. Rev. B}} \textbf{\bibinfo{volume}{89}},
  \bibinfo{pages}{180409} (\bibinfo{year}{2014}).

\bibitem{Puliafito2014}
\bibinfo{author}{{Puliafito}, V.}, \bibinfo{author}{{Siracusano}, G.},
  \bibinfo{author}{{Azzerboni}, B.} \& \bibinfo{author}{{Finocchio}, G.}
\newblock \bibinfo{title}{Self-modulated soliton modes excited in a nanocontact
  spin-torque oscillator}.
\newblock \emph{\bibinfo{journal}{IEEE Magn. Lett.}}
  \textbf{\bibinfo{volume}{5}}, \bibinfo{pages}{3000104}
  (\bibinfo{year}{2014}).

\bibitem{Wang2017}
\bibinfo{author}{Wang, C.}, \bibinfo{author}{Xiao, D.}, \bibinfo{author}{Zhou,
  Y.}, \bibinfo{author}{{\AA}kerman, J.} \& \bibinfo{author}{Liu, Y.}
\newblock \bibinfo{title}{{Phase-locking of multiple magnetic droplets by a
  microwave magnetic field}}.
\newblock \emph{\bibinfo{journal}{AIP Adv.}} \textbf{\bibinfo{volume}{7}},
  \bibinfo{pages}{56019} (\bibinfo{year}{2017}).

\bibitem{Mohseni2018}
\bibinfo{author}{Mohseni, M.} \emph{et~al.}
\newblock \bibinfo{title}{{Magnetic droplet soliton nucleation in oblique
  fields}}.
\newblock \emph{\bibinfo{journal}{Phys. Rev. B}} \textbf{\bibinfo{volume}{97}},
  \bibinfo{pages}{184402} (\bibinfo{year}{2018}).

\bibitem{Mohseni2020PRApplied}
\bibinfo{author}{Mohseni, M.} \emph{et~al.}
\newblock \bibinfo{title}{{Propagating Magnetic Droplet Solitons as Moveable
  Nanoscale Spin-Wave Sources with Tunable Direction of Emission}}.
\newblock \emph{\bibinfo{journal}{Phys. Rev. Appl.}}
  \textbf{\bibinfo{volume}{13}}, \bibinfo{pages}{24040} (\bibinfo{year}{2020}).

\bibitem{Sisodia2021}
\bibinfo{author}{Sisodia, N.}, \bibinfo{author}{Muduli, P.~K.},
  \bibinfo{author}{Papanicolaou, N.} \& \bibinfo{author}{Komineas, S.}
\newblock \bibinfo{title}{{Chiral droplets and current-driven motion in
  ferromagnets}}.
\newblock \emph{\bibinfo{journal}{Phys. Rev. B}}
  \textbf{\bibinfo{volume}{103}}, \bibinfo{pages}{24431}
  (\bibinfo{year}{2021}).

\bibitem{Yazdi2021}
\bibinfo{author}{Yazdi, H.~F.}, \bibinfo{author}{Ghasemi, G.},
  \bibinfo{author}{Mohseni, M.} \& \bibinfo{author}{Mohseni, M.}
\newblock \bibinfo{title}{{Tuning the dynamics of magnetic droplet solitons
  using dipolar interactions}}.
\newblock \emph{\bibinfo{journal}{Phys. Rev. B}}
  \textbf{\bibinfo{volume}{103}}, \bibinfo{pages}{24441}
  (\bibinfo{year}{2021}).

\bibitem{Backes2015}
\bibinfo{author}{Backes, D.} \emph{et~al.}
\newblock \bibinfo{title}{{Direct Observation of a Localized Magnetic Soliton
  in a Spin-Transfer Nanocontact}}.
\newblock \emph{\bibinfo{journal}{Phys. Rev. Lett.}}
  \textbf{\bibinfo{volume}{115}}, \bibinfo{pages}{127205}
  (\bibinfo{year}{2015}).

\bibitem{Carpentieri2015}
\bibinfo{author}{Carpentieri, M.}, \bibinfo{author}{Tomasello, R.},
  \bibinfo{author}{Zivieri, R.} \& \bibinfo{author}{Finocchio, G.}
\newblock \bibinfo{title}{{Topological, non-topological and instanton droplets
  driven by spin-transfer torque in materials with perpendicular magnetic
  anisotropy and Dzyaloshinskii-Moriya Interaction}}.
\newblock \emph{\bibinfo{journal}{Sci. Rep.}} \textbf{\bibinfo{volume}{5}},
  \bibinfo{pages}{16184} (\bibinfo{year}{2015}).

\bibitem{Hang2018}
\bibinfo{author}{Hang, J.}, \bibinfo{author}{Hahn, C.},
  \bibinfo{author}{Statuto, N.}, \bibinfo{author}{Maci{\`{a}}, F.} \&
  \bibinfo{author}{Kent, A.~D.}
\newblock \bibinfo{title}{{Generation and annihilation time of magnetic droplet
  solitons}}.
\newblock \emph{\bibinfo{journal}{Sci. Rep.}} \textbf{\bibinfo{volume}{8}},
  \bibinfo{pages}{6847} (\bibinfo{year}{2018}).

\bibitem{Jiang2018IEEEML}
\bibinfo{author}{Jiang, S.} \emph{et~al.}
\newblock \bibinfo{title}{{Impact of the Oersted Field on Droplet Nucleation
  Boundaries}}.
\newblock \emph{\bibinfo{journal}{IEEE Magn. Lett.}}
  \textbf{\bibinfo{volume}{9}}, \bibinfo{pages}{3104304}
  (\bibinfo{year}{2018}).

\bibitem{Shi2020}
\bibinfo{author}{Shi, K.} \emph{et~al.}
\newblock \bibinfo{title}{{Observation of Magnetic Droplets in Magnetic Tunnel
  Junctions}}.
\newblock \emph{\bibinfo{journal}{arXiv:}} \bibinfo{pages}{2012.05596}
  (\bibinfo{year}{2020}).

\bibitem{Nielsen1976}
\bibinfo{author}{{Nielsen}, J.}
\newblock \bibinfo{title}{Bubble domain memory materials}.
\newblock \emph{\bibinfo{journal}{IEEE Trans. Magn.}}
  \textbf{\bibinfo{volume}{12}}, \bibinfo{pages}{327--345}
  (\bibinfo{year}{1976}).

\bibitem{Giess1980}
\bibinfo{author}{Giess, E.~A.}
\newblock \bibinfo{title}{{Magnetic Bubble Materials}}.
\newblock \emph{\bibinfo{journal}{Science}} \textbf{\bibinfo{volume}{208}},
  \bibinfo{pages}{938--943} (\bibinfo{year}{1980}).

\bibitem{DeLeeuw1980}
\bibinfo{author}{De~Leeuw, F.}, \bibinfo{author}{Van Den~Doel, R.} \&
  \bibinfo{author}{Enz, U.}
\newblock \bibinfo{title}{{Dynamic properties of magnetic domain walls and
  magnetic bubbles}}.
\newblock \emph{\bibinfo{journal}{Rep. Prog. Phys.}}
  \textbf{\bibinfo{volume}{43}}, \bibinfo{pages}{689} (\bibinfo{year}{1980}).

\bibitem{Komineas1996}
\bibinfo{author}{Komineas, S.} \& \bibinfo{author}{Papanicolaou, N.}
\newblock \bibinfo{title}{{Topology and dynamics in ferromagnetic media}}.
\newblock \emph{\bibinfo{journal}{Physica D}} \textbf{\bibinfo{volume}{99}},
  \bibinfo{pages}{81--107} (\bibinfo{year}{1996}).

\bibitem{Moutafis2009}
\bibinfo{author}{Moutafis, C.}, \bibinfo{author}{Komineas, S.} \&
  \bibinfo{author}{Bland, J. A.~C.}
\newblock \bibinfo{title}{{Dynamics and switching processes for magnetic
  bubbles in nanoelements}}.
\newblock \emph{\bibinfo{journal}{Phys. Rev. B}} \textbf{\bibinfo{volume}{79}},
  \bibinfo{pages}{224429} (\bibinfo{year}{2009}).

\bibitem{Kim2003PRL}
\bibinfo{author}{Kim, D.-H.}, \bibinfo{author}{Choe, S.-B.} \&
  \bibinfo{author}{Shin, S.-C.}
\newblock \bibinfo{title}{{Direct Observation of Barkhausen Avalanche in Co
  Thin Films}}.
\newblock \emph{\bibinfo{journal}{Phys. Rev. Lett.}}
  \textbf{\bibinfo{volume}{90}}, \bibinfo{pages}{087203}
  (\bibinfo{year}{2003}).

\bibitem{Balk2014PRB}
\bibinfo{author}{Balk, A.~L.}, \bibinfo{author}{Stiles, M.~D.} \&
  \bibinfo{author}{Unguris, J.}
\newblock \bibinfo{title}{{Critical behavior of zero-field magnetic
  fluctuations in perpendicularly magnetized thin films}}.
\newblock \emph{\bibinfo{journal}{Phys. Rev. B}} \textbf{\bibinfo{volume}{90}},
  \bibinfo{pages}{184404} (\bibinfo{year}{2014}).

\bibitem{Herranen2019PRL}
\bibinfo{author}{Herranen, T.} \& \bibinfo{author}{Laurson, L.}
\newblock \bibinfo{title}{{Barkhausen Noise from Precessional Domain Wall
  Motion}}.
\newblock \emph{\bibinfo{journal}{Phys. Rev. Lett.}}
  \textbf{\bibinfo{volume}{122}}, \bibinfo{pages}{117205}
  (\bibinfo{year}{2019}).

\bibitem{Albert2020}
\bibinfo{author}{Albert, J.}, \bibinfo{author}{Maci{\`{a}}, F.} \&
  \bibinfo{author}{Hern{\`{a}}ndez, J.~M.}
\newblock \bibinfo{title}{{Effect of the Zhang-Li torque on spin-torque
  nano-oscillators}}.
\newblock \emph{\bibinfo{journal}{Phys. Rev. B}}
  \textbf{\bibinfo{volume}{102}}, \bibinfo{pages}{184421}
  (\bibinfo{year}{2020}).

\bibitem{Graefe2019}
\bibinfo{author}{{J. Gr{\"{a}}fe, M. Weigand, B. Van Waeyenberge, A. Gangwar,
  F. Gro{\ss}, F. Lisiecki, J. Rychly, H. Stoll, N. Tr{\"{a}}ger, J.
  F{\"{o}}rster, F. Stobiecki, J. Dubowik, J. Klos, M. Krawczyk, C. H. Back, E.
  J. Goering, G. Sch{\"{u}}tz, H.-J. M. Drouhin, J.-E. Wegrowe, and M.
  Razeghi}}.
\newblock \bibinfo{title}{{Visualizing nanoscale spin waves using MAXYMUS}}.
\newblock In \emph{\bibinfo{booktitle}{Proc. SPIE}}, vol.
  \bibinfo{volume}{11090}, \bibinfo{pages}{1109025}
  (\bibinfo{publisher}{Spintronics XII}, \bibinfo{year}{2019}).

\bibitem{Nolle2012}
\bibinfo{author}{Nolle, D.} \emph{et~al.}
\newblock \bibinfo{title}{{Note: Unique characterization possibilities in the
  ultra high vacuum scanning transmission x-ray microscope (UHV-STXM) "MAXYMUS"
  using a rotatable permanent magnetic field up to 0.22 T}}.
\newblock \emph{\bibinfo{journal}{Rev. Sci. Instrum.}}
  \textbf{\bibinfo{volume}{83}}, \bibinfo{pages}{046112}
  (\bibinfo{year}{2012}).

\end{thebibliography}

\begin{methods}

\subsection{Sample Preparation} A sample stack is consisted of a Ta~($4$~nm)/ Cu~($14$~nm) / Ta~($4$~nm) / Pd~($2$~nm) seed layer and an all-perpendicular pseudo-spin valve [Co~($0.35$~nm) / Pd~($0.7$~nm)]${\times5}$ / Co~($0.35$~nm) / Cu~($5$~nm) / [Co~($0.22$~nm) / Ni~($0.68$~nm)]${\times4}$ / Co~($0.22$~nm), capped by a Cu~($2$~nm) / Pd~($2$~nm) layer, which was deposited by magnetron sputtering on Si wafer with 300~nm thick LPCVD silicon nitride layer. Using a conventional photo-lithography and metal-etching techniques, 8 $\mu$m $\times$ 16 $\mu$m mesas were fabricated on above stack wafer and all mesas were insulated by a 30-nm-thick SiO$_2$ layer deposited by using chemical vapor deposition (CVD). To pattern nanocontacts (NCs) on the top of each mesa having different diameters from 50 to 150~nm, electron beam lithography was used. SiO$_2$ layer was then etched through by the reactive ion etching (RIE) technique to open NCs. The NC-STO device fabrication was completed by the deposition of Cu $200$~nm / Au $100$~nm top electrode and lift-off processing. For STXM measurements, Si was removed from backside using highly selective RIE process and leave only SiN membrane to allow X-ray transmission underneath NC-STOs. (See, Figure~\ref{figure1}(c)) For magnetic and electrical charaterization of NC-STOs, same stack was prepared on Si thermally oxidized Si wafer and then similar fabrication processing were done except a deep etching for a membrane structure.
\par

\subsection{Magnetic and Electrical Characterization} the magnetization hysteresis loops was measured using Alternating Gradient Magnetometry (AGM) with the unpatterned material stacks. \textit{dc} and microwave measurements of the fabricated STOs were carried out using our custom-built setup, where magnetic field strength, polarity, and angle can be controlled. A magnetic field between -0.5 to +0.5 can be manipulated using electromagnet. The device is connected using GSG probe to a \textit{dc}-current source (Keithley 6221), a nanovoltmeter (Keithley 2182A), and a spectrum analyzer (R \& S FSQ26). A 0--40GHz bias-tee is used to separate the bias input and the generated microwave signal. The microwave sinalg is amplified by a low-noise amplifier (operational range: 0.1--26.5 GHz) before being sent to the spectrum analyzer.\par

\subsection{Scanning transmission x-ray microscopy} The STXM measurements were performed at the BESSY II synchrotron, using the MPI~IS operated MAXYMUS end station at the UE46-PGM2 beam line.\cite{Graefe2019} The out-of-plane component of the magnetization was probed using circularly polarized light at normal incidence. The applied field, with a maximum value of 300~mT, was generated by a set of four rotatable permanent magnets\cite{Nolle2012}. An optimal XMCD contrast was achieved by setting the photon energy to the Ni~$L_{3}$ edge, which resulted in clear images. The size of each pixel is 10$\times$10~$nm^2$, while the nominal resolution of the focusing plate is 18~nm. 
 
\end{methods}

\begin{addendum}
\item This work was supported by the Swedish Research Council (VR; 2017-06711 and 2019-04229). Helmholtz Zentrum Berlin is acknowledged for allocating beam time at the BESSY II synchrotron radiation facility. M. W., E.G., G.S. and J.G. acknowledge the financial support by the Federal Ministry of Education and Research of Germany in the frame work of DynaMAX (Project No. 05K18EYA). This work was supported by the National Research Foundation of Korea(NRF) grant funded by the Korea government(MSIT) (No. 2020R1F1A1049642)

\item These authors contributed equally: Martina Ahlberg and Sunjae Chung.
\item [Contributions] M.A., S.C., and J.\AA{}. conceived the project, S.C. S. J. and T.Q.L. performed the electrical measurements. S.C., T.Q.L., S.J., and A.H. fabricated the devices. M.A., S. J., J.G., M.W., F.G. and I.B. carried out the STXM measurements. 
J.\AA{}. coordinated the project. All authors analyzed the results and co-wrote the manuscript.
\item[Competing Interests] The authors declare that they have no competing financial interests.
\item[Correspondence] Correspondence and requests for materials should be addressed to S. Chung (email: sjchung76@knue.ac.kr) and J. \AA{}kerman (email: johan.akerman@physics.gu.se).
\end{addendum}

\clearpage

\beginsupplement
\subsection{\textbf{Supplementary information.} } \label{sup1}

\begin{figure}
  \begin{center}
  \includegraphics[width=6in]{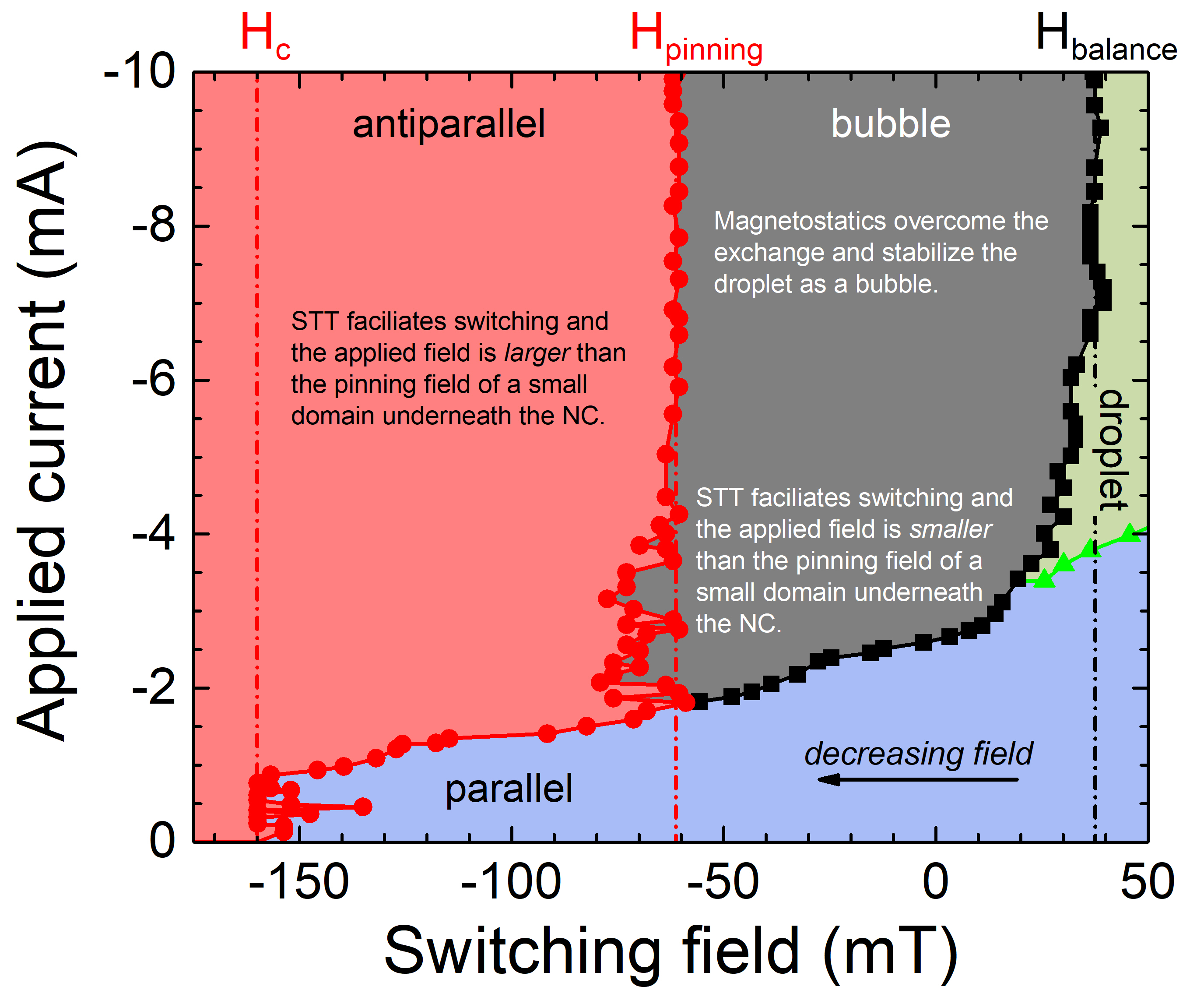}
    \caption{\textbf{Supplementary Fig.1. }
    A zoom-in of the phase diagram in Fig.~3(a) of the main text. The color code represents the different states: droplet (green), bubble (gray) and antiparallel (red). A droplet is nucleated at high currents and fields. Below a certain positive field ($H_{\mathrm{balance}}$) the droplet is stabilized as a static bubble due to magnetostatic effects\cite{Hoefer2010}. The bubble is pinned below the nanocontact until the negative field is high enough to let the bubble domain expand throughout the film at $H_{\mathrm{pinning}}$. At low currents the magnetic switching is only governed by the coercive field ($H_{\mathrm{c}}$) of the free layer.}
    \label{figureS1}
    \end{center}
\end{figure}
\clearpage

\end{document}